\documentclass[aps,apl]{revtex4}

\usepackage{amsmath,amssymb}
\input epsf
\usepackage{verbatim}
\usepackage{graphicx}
\usepackage{bm}
\usepackage{natbib}
\def\pmb#1{\setbox0=\hbox{#1}
\kern-.025em\copy0\kern-\wd0 \kern-.05em\copy0\kern-\wd0
\kern-.025em\raise.0433em\box0}

\newcommand{\C}{\mathbb C}

\newcommand{\beq}{\begin{equation}}
\newcommand{\eeq}{\end{equation}}
\newcommand{\ba}{\begin{eqnarray}}
\newcommand{\ea}{\end{eqnarray}}

\input epsf

\usepackage[english]{babel}
\begin{document}
%\doublespace

\title[]{Controlling solid elastic waves with spherical cloaks
% lying atop a
 % Faqir's bed
 % of nails
}
\author{Andre Diatta$^{(1)}$
, Sebastien Guenneau$^{(2)}$
}
\affiliation{ Aix$-$Marseille Universit\'e, CNRS, Centrale Marseille, Institut Fresnel, UMR CNRS 7249, 13013 Marseille, France\\
$(1)$ andre.diatta@fresnel.fr, ~~ $(2)$ sebastien.guenneau@fresnel.fr }
%\footnote{(1) andre.diatta@fresnel.fr, ~~ (2) sebastien.guenneau@fresnel.fr}
\begin{abstract}
We propose a cloak for coupled shear and pressure waves in solids. Its elastic properties are deduced from a geometric transform that retains the form of Navier equations. The spherical shell is made of an anisotropic and  heterogeneous medium described by an elasticity tensor ${\C}'$ (without the minor symmetries) which has 21 non-zero spatially varying coefficients in spherical coordinates.  Although some entries of ${\C}'$, e.g. some with a radial subscript, and the density (a scalar radial function)  vanish on the inner boundary of the cloak, this metamaterial exhibits less singularities than its cylindrical counterpart studied in [M. Brun, S.  Guenneau, A.B. Movchan, Appl. Phys. Lett. 94, 061903 (2009).]
 In the latter work, ${\C}'$ suffered some infinite entries, unlike in our case. Finite element computations confirm that elastic waves are smoothly detoured around a spherical void  without reflection. % Finite element computations reveal that the displacement field vanishes inside the inner core of the shell except for a countable set of eigenfrequencies corresponding to eigenmodes of the stress-free cavity.


\end{abstract}
%\label{firstpage}
\maketitle

\section{introduction}
In 2006, Pendry, Schurig and Smith \cite{pendry}, and Leonhardt \cite{leonhardt} independently proposed some design of invisibility cloaks for light. The same year, Milton, Briane and Willis laid the foundations of transformational physics \cite{milton2006}. All these works generated a lot of  interest in the metamaterials community, notably in the design of electromagnetic \cite{13}, water wave \cite{prl2008} and acoustic \cite{cummer-schurigt2007,sanchez2007,chen2007,norris2008,cummer-pendry-prl-mie} cloaks. The latter fuelled the interest in acoustic metamaterials \cite{21}, which enable a markedly enhanced control of pressure waves, including lensing,  via artificial anisotropy \cite{20}.

In 2009, Brun et al. in \cite{brunapl},  
%\cite{23},
  discussed the design of a cylindrical cloak for in-plane elastic waves with an asymmetric elasticity tensor, which was a modified version of the Willis' type \cite{willis1981} transformed equations derived in \cite{milton2006}. In 2011, these two works were encompassed by Norris and Shuvalov in a more general elasticity framework \cite{norris2011}.

In parallel to the developments of metamaterials for bulk elastic waves, some theoretical \cite{farhat2009,farhat2012} and experimental \cite{stenger2012} progress was made in the control of flexural elastic waves in thin plates. In the case of thin plates, the transformed
governing equations (e.g. Kirchhoff) have a simpler structure which helps engineer structured cloaks.

In the present letter, we investigate spherical cloaks
for solid elastic waves using a radially symmetric linear
geometric transform.
We discuss their underlying mechanism and illustrate
the theory using a finite element approach which is adequate to solve the Navier
equations in transformed anisotropic heterogeneous media
with asymmetric elasticity tensors.

\section{Transformed equation and elastic properties of cloak}

\subsection{The equations of motion}

The propagation of
elastic waves is governed by the Navier equations. Assuming
time harmonic $\exp(-i\omega t)$ dependence, with $\omega$ as the
angular wave frequency and $t$ the time variable, allows us to work directly in the spectral domain.
Such dependence is assumed henceforth and suppressed, leading to
\begin{equation}
\nabla\cdot{\boldsymbol{\sigma}}=-i\omega {\bf p} \; , \; \; \;  \;  \;
\boldsymbol{\sigma}={\C}:\nabla{\bf u} \; ,  \; \; \; \;  \; 
{\bf p}=-i\omega\rho {\bf u} \; , \;
\label{navier}
\end{equation}
where $\rho$ is the density and $\sigma$ the stress tensor of the (possibly heterogeneous
isotropic) elastic medium, ${\bf u}=(u_r,u_\theta,u_\phi)$ and
$\C$ are respectively the
three-component displacement field and the rank-four
(symmetric) elasticity tensor expressed
in a spherical coordinate basis ${\bf x}=(r,\theta,\phi)$.

\subsection{The transformed equations of motion}

Let us consider the coordinate change ${\bf x}\longmapsto {\bf x}'$,
where ${\bf x'}=(r',\theta',\phi')$ are stretched spherical coordinates.
This  leads to
%a transformed displacement ${\bf u}'$ such that ${\bf u}={\bf A}^T{\bf u}'$ and
a transformed equation \cite{norris2011}
 
\begin{equation}
\begin{array}{ll}
%\nabla\cdot{\bf C}:\nabla{\bf u}+\rho\omega^2 {\bf u}={\bf 0} \; ,
\nabla'\cdot{\boldsymbol{\sigma}'}=-i\omega {\bf p}' \; , \;  \; \; \;  \;
\boldsymbol{\sigma}'={\C}':\nabla'{\bf u'}+{\bf S}{\bf u'} \; , \; \\
{\bf p}'={\bf D}\nabla'{\bf u'}-i\omega{\boldsymbol\rho}{\bf u}' \; , \;  \; \; \;  \; 
{\bf u'}={\bf A}^{-T}{\bf u} \; ,
\end{array} 
\label{navierwillis}
\end{equation}
in general. Here  {\bf S} and {\bf D} are third order tensors possibly encompassing a $i\omega$ dependence, $\nabla'$ is the gradient in transformed coordinates ${\bf x'}$  and
${\bf u}'({\bf x'})={\bf u'}(r',\theta',\phi')$ is a transformed displacement
in stretched spherical coordinates.

Note that the transformed stress  $\sigma'$ is generally not symmetric.
Note also that in general ${\bf A}$ is a matrix field.

One possibility to preserve
the symmetry of the stress tensor, is to
assume that  ${\bf A}$ is a multiple ${\bf A}=\xi\partial{\bf x}'/\partial{\bf x}$, of the Jacobian matrix $\partial{\bf x}'/\partial{\bf x}$ of the transformation, 
where $\xi$ is a non-zero scalar, in which case one obtains a  Willis-type equation \cite{willis1981,milton2006}.
However, there is one special case for which ${\bf u}'={\bf u}$ when
${\bf A}$ is the identity matrix ${\bf I}$, which leads to  \cite{brunapl}

\begin{equation}
\nabla'\cdot{\C}':\nabla'{\bf u'}+\rho'\omega^2 {\bf u'}={\bf 0} \;
,\label{snavier}
\end{equation}
where the body force is assumed to be zero, the elasticity tensor
${\C}'$ does not have the minor symmetries (Cosserat Material)  and the stretched density
$\rho'$ is
a scalar field. This equation is derived from (\ref{navier}) by
noting that ${\bf S}={\bf D}={\bf 0}$  when
${\bf A}={\bf I}$,\;\cite{norris2011}. 
In the sequel, we work in the framework of (\ref{snavier}).

\begin{figure}[h!]
\resizebox{175mm}{!}{\includegraphics{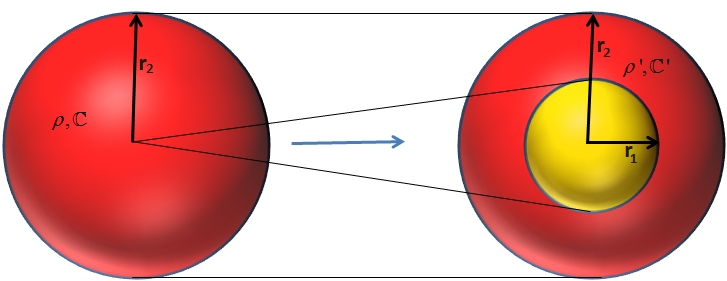}}
%\vspace{-3mm}
\caption{Geometric transform (\ref{PTransform}) from ${\bf x}=(r,\theta,\phi)$
(left)
to ${\bf x}'=(r',\theta',\phi')$ (right), where $r_1$ and $r_2$ are the inner and outer
radii of the spherical cloak, respectively. A sphere (left) with an
isotropic homogeneous symmetric elastic constitutive tensor $\C$ and homogeneous scalar density $\rho,$
is mapped onto a hollow sphere (right) with a heterogeneous asymmetric elastic constitutive tensor $\C'$
and a heterogeneous scalar density $\rho'$.
}
\label{3d_cloak_sphere_elastic1}
\end{figure}

Let us consider the geometric transform
\begin{equation}
r'=r_1+\frac{r_2-r_1}{r_2}r \; , \; \theta'=\theta
\; , \; \phi'=\phi \; , \label{PTransform}
\end{equation}
which maps a sphere of radius $0<r\leq r_2$ onto on shell
$r_1<r'\leq r_2$, see Figure \ref{3d_cloak_sphere_elastic1}.

Design of transformation-based Cosserat elastic cloaks has been
first discussed in the cylindrical case in \cite{brunapl}  where it only involved a tensor
$\C'$ with $8$ non-vanishing coefficients, whereas in the present spherical
case, we need to consider a tensor $\C'$ with $21$ 
non-vanishing coefficients. Moreover, the displacement field
has three components in our case.

By application of transformation (\ref{PTransform}), in the region
$ 0<r\leq r_2,$ the Navier equations (\ref{navier}) are mapped into the
equations (\ref{snavier}) with
%\begin{equation}
%\rho'= {\left(\frac{r'-r_1}{r'}\right)}^2{\left(\frac{r_2}{r_2-r_1}\right)}^3\,\rho \; , \label{srho}
%\end{equation}
\begin{equation}
\rho'= ab^2(r') \,\rho \; , \label{srho}
\end{equation}
where $a=\frac{r_2}{r_2-r_1}$ and $b(r')=\frac{(r'-r_1)}{r'}a$,
and the elasticity tensor ${\bf \C}'$ has 21 non-zero spherical components, namely
\begin{eqnarray}
C_{r'r'r'r'}'= (\lambda+2\mu)\frac{b^2(r')}{a},  %\nonumber \\
C_{\theta\theta\theta\theta}'=C_{\phi \phi\phi\phi}'=  (\lambda+2\mu)a,
\nonumber \\
C_{r'r'\theta\theta}'=  C_{\theta\theta r'r'}'=C_{r'r'\phi\phi}'= C_{\phi \phi r'r'}'=\lambda b(r'), \nonumber \\
C_{\theta\theta\phi\phi}'=  C_{\phi\phi\theta\theta}'=\lambda a, ~
C_{r'\theta r'\theta}'=  C_{r'\phi r'\phi}'=\frac{b^2(r')}{a}\mu, \nonumber \\
C_{\theta r'\theta r'}'=  C_{\theta\phi \theta\phi}'=C_{\phi r' \phi r' }'=  C_{\phi\theta\phi \theta}'=a\mu, \nonumber
\\
C_{r'\theta\theta r'}'=  C_{\theta r'r'\theta}=  C_{r'\phi\phi r'}'= C_{\phi r'r' \phi }'=b(r')\mu, \nonumber 
\\
C_{\theta\phi \phi\theta}'=C_{\phi\theta\theta \phi}'=a\mu \;  . \;
\label{sc}
\end{eqnarray}

\subsection{Singularity at the cloak's inner boundary}

As discussed above, one should note that the minor symmetries are broken and $b(r_1)=0$ which means
$C_{\theta\theta\theta\theta}'/C_{r'r'r'r'}'=C_{\phi \phi\phi\phi}'/C_{r'r'r'r'}'$ is infinite
at the inner boundary of the cloak (infinite anisotropy). Moreover, the off-diagonal components
are constant or vanish at the boundary $r'=r_1$.

 Physically, this means that shear and pressure waves propagate much faster in the azimuthal and elevation directions than in the radial direction on the surface of the inner boundary. This should result in a vanishing phase shift between an elastic wave propagating in an isotropic homogeneous elastic medium, and another one propagating around the concealed region.

One should also note that there are no infinite entries within the elasticity tensor. This is a feature of the 3D  elastodynamic cloak which  is less singular than its 2D counterpart (the cylindrical elastic
cloak in \cite{brunapl}  had a vanishing density on the inner boundary of the cloak as well, but some entries of the elasticity tensor
going to infinity at the inner boundary).

\subsection{Impedance matching at the cloak's outer boundary}

Let us now note that when $r=r_2$, the geometric transform
(\ref{PTransform}) leads to $r'=r_2$. In this case, the transformed
density is $\rho'=a\rho$ as $b(r_2)=1$ and the diagonal
components of the transformed
elasticity tensor reduce to
%\begin{equation}
$
C'_{r'r'r'r'}=(\lambda+2\mu)/a \; ,
C'_{\theta\theta\theta\theta}=
C'_{\phi\phi\phi\phi}
=(\lambda+2\mu)a \; ,
$
%\end{equation}
hence $C'_{r'r'r'r'}=C_{rrrr}/a$ and
$C'_{\theta\theta\theta\theta}=a C_{\theta\theta\theta\theta}$
and
$C'_{\phi\phi\phi\phi}=a C_{\phi\phi\phi\phi}$
on the outer boundary of the cloak.
This expresses the fact that a stretch along the radial direction
is compensated by a contraction along the azimuthal
and elevation directions in such a way that elastic media (cloak and surrounding isotropic elastic medium) are
impedance-matched at $r'=r_2$.
The cloak's outer boundary therefore behaves in many ways as an
impedance matched `thin' elastic layer.
However, we note that the components of ${\C}'$ pose no limitations
on the applied frequency $\omega$ from low to high frequency, unlike for the
case of coated cylinders studied back in 1998 in the context of elastic
neutrality by Bigoni et al. \cite{bigoni98}.

The fact that ${\C}'$ does not depend on $\omega,$ i.e. the cloak consists
of a non-dispersive elastic medium, makes it work at all frequencies, but
one should keep in mind that any structured medium designed to approximate
the ideal cloak's parameters (e.g. via homogenization) would necessarily
involve some dispersion, and thus limit the interval of frequencies over
which the cloak can work. Such a feature has been already observed
in \cite{farhat2012,stenger2012}, for cloaking of flexural waves in thin-elastic plates.

\section{Numerical illustration}

\begin{figure}[h!]
\resizebox{175mm}{!}{\includegraphics{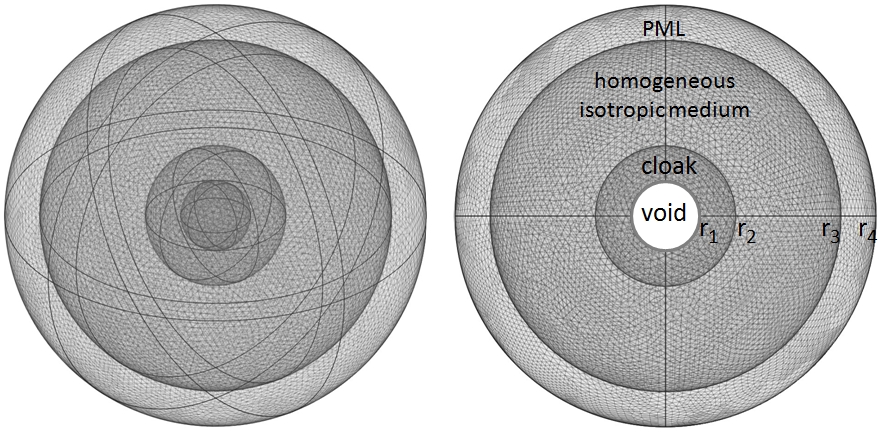}}%{mesh10022014a.jpg}}
%\vspace{-3mm}
\caption{A 3D plot (left) and slice (right) of the mesh of the
computational domain;
the mesh has $1105932$ tetrahedral elements, $36492$ triangular elements,
$1128$ edge
elements and $25$ vertex elements; an anisotropic heterogeneous absorptive
(and reflectionless) perfectly matched layer (PML) occupies a spherical
shell of
inner radius $r_3=10$~m and outer radius $r_4=12$~m; the shell surrounds 
an isotropic homogeneous elastic sphere of radius $r_3$, which contains
a void surrounded by a cloak. The latter is an anisotropic heterogeneous spherical
shell of inner radius $r_1=2$~m and outer radius $r_2=4$~m.}
\label{3d_cloak_sphere_elastic_mesh}
\end{figure}

We would like now to numerically test the cloaking efficiency. For this, we implement the $3^4$ spatially varying
entries of the transformed tensor in Cartesian coordinates in the finite element package COMSOL MULTIPHYSICS.
We mesh the computational domain using 1105932 tetrahedral elements, 36492 triangular elements, 1128 edge
elements and 25 vertex elements, see Figure \ref{3d_cloak_sphere_elastic_mesh}. %This domain consists of a sphere of radius $r_3=10$~ m (with the void, the cloak and the point force located within an isotropic homogeneous elastic medium), and a surrounding spherical shell of inner radius $r_3$ and outer radius $r_4=12$~m with an anisotropic heterogeneous absorptive (and reflectionless) perfectly matched layer (PML).
This domain consists of  an isotropic homogeneous elastic medium within a sphere of radius $r_3=10$~m, containing  a void surrounded by the cloak (a heterogeneous, anisotropic elastic metamaterial in a spherical shell of inner radius $r_1=2$~m and outer radius $r_2=4$~m ) 
and  a point force located at $(4 ~ m,-7 ~m,0~m),$ where $(0,0,0)$ is the center of the cloak. 
The sphere of radius $r_3=10$~m is itself surrounded by a spherical shell of inner radius $r_3$ and outer radius $r_4=12$~m, which is filled with an anisotropic heterogeneous aborptive medium acting as a (reflectionless) perfectly matched layer (PML).
This elastic PML is deduced from a geometric transform

\begin{equation}
r''=r_3+\displaystyle\int_{r_3}^{r}s_r(v)dv \; , \; \theta''=\theta
\; , \; \phi''=\phi \; , \label{PML}
\end{equation}
where we consider the radial function $s_r(r)=1-i.$ This transform leads, in the same way as (4) did for $\C'$ and $\rho'$, to a heterogeneous anisotropic asymmetric elasticity tensor $\C''$ and a scalar (homogeneous) density $\rho'',$ whose expressions are  obtained by taking $a=1-i$, $b(r'')=1$ in  (\ref{srho})-(\ref{sc}). Since the elastic waves are damped inside the PML and reach the outer boundary of the shell with a vanishing amplitude, we set either clamped or traction free boundary conditions at $r''=r_4$ (we have checked this does not affect the numerical result).

\begin{figure}[h!]
\resizebox{175mm}{!}{\includegraphics{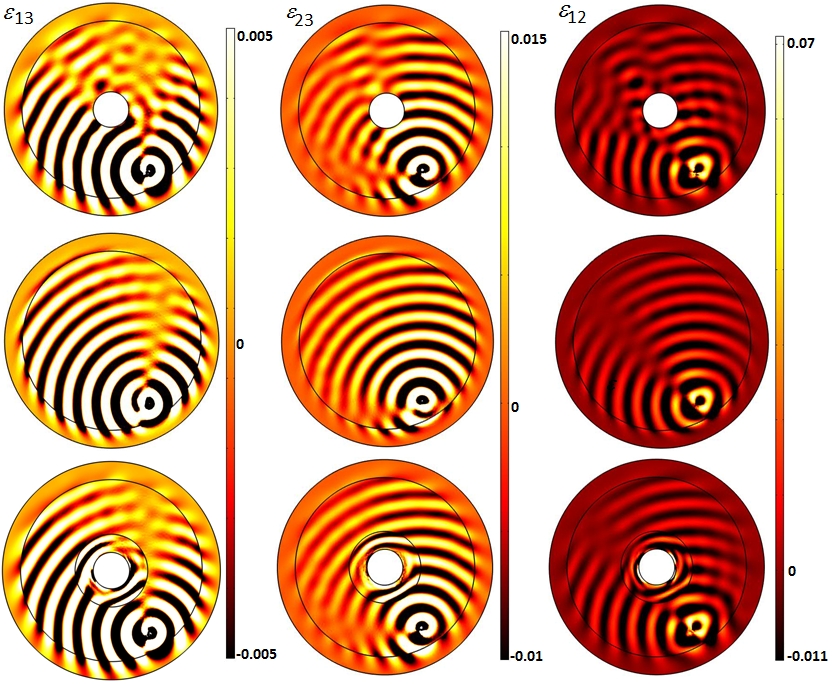}}
%\vspace{-3mm}
\caption{Deformation of an isotropic elastic medium (middle panel), a stress-free spherical obstacle of radius $2$ m (upper panel) and the same void surrounded by an elastic cloak (lower panel) of inner radius $r_1=2$\; m and outer radius $r_2=4$\; m, subjected to a concentrated load of polarization $(1, 1, 1)$ with frequency  $3Hz$ located at a distance $8.062$ m from the origin. Deformation of components $\varepsilon_{13} =\frac{1}{2}[\frac{\partial u_1}{\partial x_3} + \frac{\partial u_3}{\partial x_1}]$
(left column),   $\varepsilon_{23} =\frac{1}{2}[\frac{\partial u_2}{\partial x_3} + \frac{\partial u_3}{\partial x_2}]$ (middle column) and  $\varepsilon_{12} =\frac{1}{2}[\frac{\partial u_1}{\partial x_2} + \frac{\partial u_2}{\partial x_1}]$ (right column) of the strain tensor $\varepsilon$.
}
\label{3d_cloak_sphere_elastic2}
\end{figure}

\medskip

\begin{figure}[h!]
\resizebox{175mm}{!}{\includegraphics{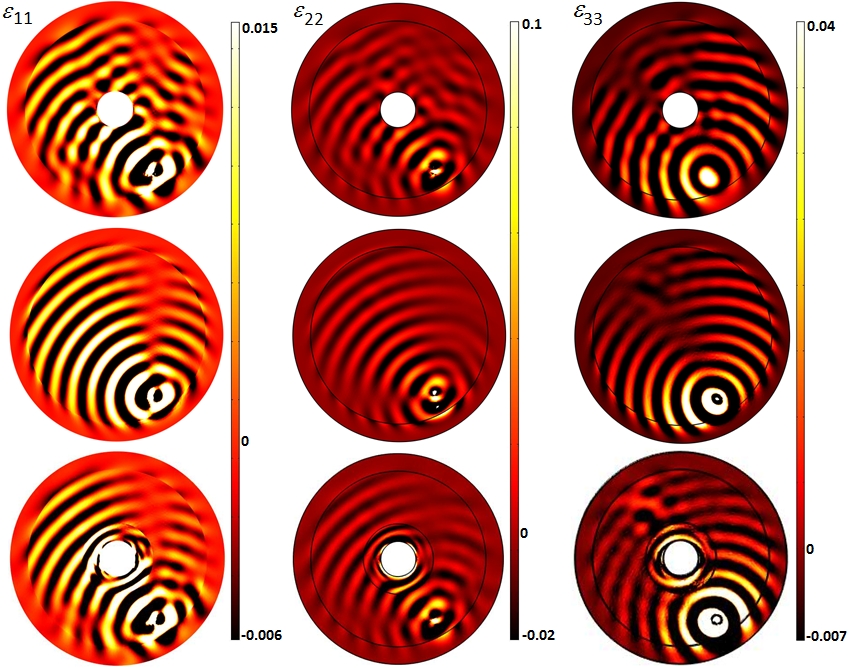}}
%\vspace{-3mm}
\caption{
Deformation of an isotropic elastic medium (middle panel), a stress-free spherical obstacle of radius $2$ m (upper panel) and the same void surrounded by an elastic cloak (lower panel) of inner radius $r_1=2$~m and outer radius $r_2=4$~m, subjected   to a concentrated load of polarization $(1, 1, 1)$ with frequency  $3~Hz$ located at a distance $8.062$~m from the origin.  Deformation of components $\varepsilon_{11}=\frac{\partial u_1}{\partial x_1}$
(left column), $\varepsilon_{22}=\frac{\partial u_2}{\partial x_2}$  (middle column) and
$\varepsilon_{33}=\frac{\partial u_3}{\partial x_3}$ (right column) of the strain tensor $\varepsilon$.
}
\label{3d_cloak_sphere_elastic3}
\end{figure}

\begin{figure}[h!]
\resizebox{175mm}{!}{\includegraphics{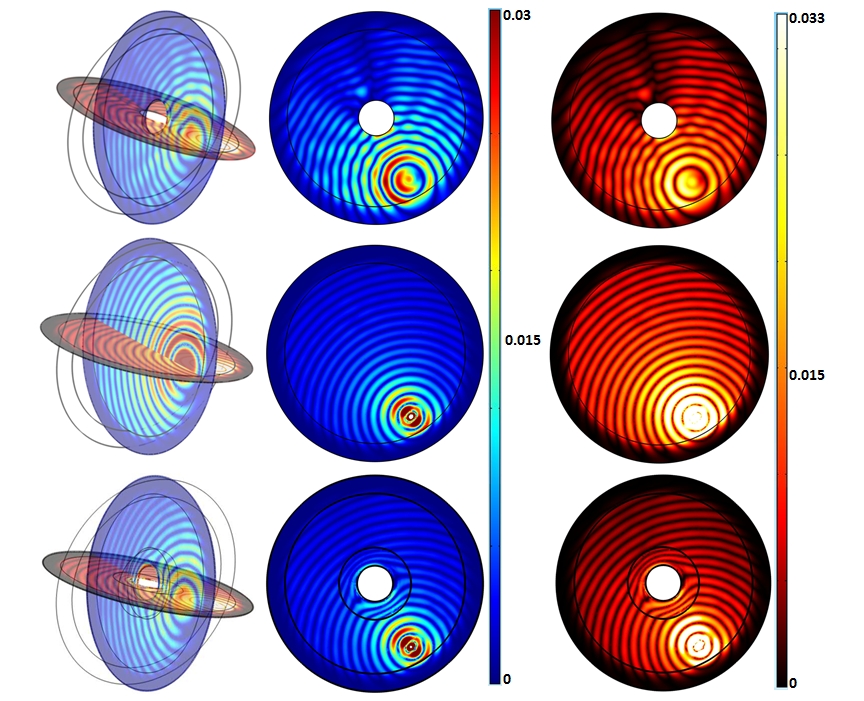}}
%\vspace{-3mm}
\caption{Slices of the 3D plot of the magnitude  $\sqrt{u_1^2+u_2^2+u_3^2}$  of the elastic  displacement field, for the point force in presence of the void (upper panel), the void surrounded by the cloak (lower panel) and in the isotropic elastic medium (middle panel). The middle and the left columns are the yz and xy slices, respectively.  Parameters are as in Figures  \ref{3d_cloak_sphere_elastic2}-\ref{3d_cloak_sphere_elastic3}.
}
\label{3d_cloak_sphere_elastic4}
\end{figure}

One should keep in mind that any numerical implementation in a finite element package requires a Cartesian coordinate system. In our case, we used COMSOL where the transformed elastic tensors $\C'$ for the cloak and $\C''$ for the PML had up to $81$ non-vanishing spatially varying entries. A good way to detect any flaw in the numerical implementation is to compare the gradient of the solution to the
problem for a time harmonic point source in an isotropic homogeneous medium (supplied with spherical elastic PML),
see middle panel in Figures \ref{3d_cloak_sphere_elastic2}-\ref{3d_cloak_sphere_elastic3}, and
the solution to the same problem when we have a cloak surrounding certain stress-free spherical region
(e.g. a void in soil) which amounts to assuming that
$\boldsymbol{\sigma}'\cdot{\bf n}=({\C}':\nabla'{\bf u})\cdot{\bf n}={\bf 0}$ where ${\bf n}$ is the
outward unit normal to the boundary of the void, see
lower panel in Figures \ref{3d_cloak_sphere_elastic2}-\ref{3d_cloak_sphere_elastic3}.
 One can see that the deformation of the
elastic medium outside the cloak is nearly identical to that of
the isotropic homogeneous elastic medium (we use the normalized density $\rho=1$
and Lam\'e parameters $\mu=1$ and $\lambda=2.3$, for respectively the shear modulus and the compressibility).
By comparison, the deformation of the elastic medium is clearly visible for the void when it is not surrounded by the cloak, see upper panel in Figures  \ref{3d_cloak_sphere_elastic2}-\ref{3d_cloak_sphere_elastic3}.
 The small discrepancy between the middle and lower panels in Figures  \ref{3d_cloak_sphere_elastic2}-\ref{3d_cloak_sphere_elastic3} is attributed to the numerical approximation of the infinitely anisotropic tensor at the inner boundary of the cloak (one way to avoid the singularity would be to use Kohn's transfrom \cite{kohn2}).
 Plots of elastic deformation can be somewhat misleading, and we therefore add plots of magnitude of the elastic displacement field in Figure  \ref{3d_cloak_sphere_elastic4}, where it should be noticed that the shaded region behind the void in the upper panel (drop of wave amplitude and phase shift) , is almost completely removed in the lower panel thanks to the cloak. Upon inspection of the middle and lower panels, one can clearly see that the elastic field scattered by a void clothed with the cloak is virtually indistinguishable from bare isotropic elastic space.

\section{Some possible applications and conclusion}
 
Finally, we would like to stress that another aspect of the spherical elastic cloak
which we designed is its capability to protect any object placed inside the spherical void
within the cloak from incoming elastic waves. Applications in anti-earthquake devices
make this feature quite interesting to investigate, as typical frequencies of earthquakes
are from $0.1$ to $10$~Hz, which is compatible with the frequency $3$~Hz for a
void of radius $2$~m studied in the numerical illustrations of this letter. Besides from that,
one can easily scale up the void e.g. a void of radius $20$ m surrounded by a spherical
cloak of outer radius $40$~m would react in exactly the same way to a concentrated load of polarization $(1,1,1)$ with pulsating frequency $0.3$~Hz located at a distance $r= 80.62$~m from the center of the cloak.

Moreover, the material parameters of our
cloak are not frequency dependent, and we numerically checked that similar results
to those shown in the present letter occur for frequencies from $1$ Hz to $10$ Hz (we computed ten evenly spaced frequencies):
we were limited below this range by the accuracy of the spherical PML (the larger the wave wavelength,  the thicker the PML and the more absorption needed)
 and above this range
we are limited by the computational resources (numerical computation at 10 Hz required about 2 million tetrahedral elements for a converged result).
The same remark holds for a void of radius $r_1=20$~m surrounded by a cloak of outer radius $r_2=40$~m as in this case the range of frequencies where the seismic signal would be detoured around the void is from $0.1$~Hz to $1$~Hz. Therefore, our study might find some
applications in seismic metamaterials \cite{kim2012,prl2014}.

We finally note that recent advances in fabrication and characterization of elastic metamaterials \cite{kadic2012,kadic2013}
 could foster experiments in an approximate 3D elastic cloak. Of course, the metamaterial would only be able in practice to display a strong (not infinite) anisotropy on the cloak's inner boundary and it would only work throughout a finite range of frequencies. Its properties could be derived for instance from an effective medium approach in a similar way to what was proposed \cite{farhat2012} and experimentally validated [18] for elastic waves in thin plates. However, detouring solid waves (even partially) around a void is much more challenging than what was done for surface waves in plates. We hope the prospect of civil engineering applications \cite{prl2014} will fuel the research in seismic cloaks.

\section*{ Acknowledgement}
The authors acknowledge European funding through ERC Starting Grant ANAMORPHISM.


\begin{thebibliography}{99}
\bibitem{pendry}
 J.B. Pendry, D. Schurig and D.R. Smith, 
Controlling Electromagnetic Fields,
{ Science} {312} 1780 (2006).
\bibitem{leonhardt}
 U. Leonhardt,
Optical Conformal Mapping,
{ Science} 312, 1777 ( 2006).
\bibitem{milton2006}%{22}
G.W. Milton, M.  Briane,  and J.R. Willis, 
On cloaking for elasticity and physical equations with a transformation invariant form.
New J. Phys. 8, 248 (2006).
\bibitem{13}
D. Schurig, J.J. Mock, B.J. Justice, S.A. Cummer, J.B. Pendry, A.F. Starr,
D.R. Smith,
Metamaterial electromagnetic cloak at microwave frequencies.
Science 314, 977-980 (2006).
\bibitem{prl2008}
M. Farhat, S. Enoch, S. Guenneau, A.B. Movchan, Broadband Cylindrical Acoustic Cloak for Linear Surface Waves in a Fluid, 
Phys. Rev. Lett. 101, 134501 (2008).

\bibitem{cummer-schurigt2007}
S. A. Cummer and D. Schurig, One path to acoustic cloaking, New J. Phys. 9, 45 (2007).

\bibitem{sanchez2007}
D. Torrent and J. Sanchez-Dehesa,  Acoustic metamaterials for new two dimensional sonic devices, New J. Phys. 9, 323
(2007).
\bibitem{chen2007}
H. Chen and C.T. Chan, Acoustic cloaking in three dimensions using acoustic metamaterials, Appl. Phys. Lett. 91, 183518 (2007).

\bibitem{norris2008}
A. Norris, Acoustic cloaking theory,  Proc. R. Soc. London, Ser. A 464, 2411 (2008).

\bibitem{cummer-pendry-prl-mie}
A. S. Cummer, B.-I.  Popa, D. Schurig, D.R. Smith, J. Pendry, M. Rahm, and A. Starr,
Scattering Theory Derivation of a 3D Acoustic Cloaking Shell,
Phys. Rev. Lett. 100, 024301 (2008).

\bibitem{21}
R.V. Craster  and S. Guenneau,
 Acoustic Metamaterials: Negative Refraction, Imaging, Lensing and Cloaking. 
Springer-Verlag, Springer Series in Materials Science, (2013).

\bibitem{20}
J. Christensen  and F.J. Garcia de Abajo,
Anisotropic metamaterials for full control of acoustic waves.
Phys. Rev. Lett. 108, 124301 (2012).

\bibitem{brunapl}%\bibitem{23}
M. Brun, S.  Guenneau  and A.B. Movchan,
Achieving control of in-plane elastic waves.
Appl. Phys. Lett. 94, 061903 (2009).

\bibitem{kohn2}
R.V. Kohn, H. Shen, M.S. Vogelius, and M.I. Weinstein, Cloaking via change of variables in electric impedance tomography, 
Inverse Problems 24, 015016 (2008).

\bibitem{willis1981}
J.R. Willis,
Variational principles for dynamic problems for inhomogeneous elastic media
Wave Motion 3, 1-11 (1981).

\bibitem{norris2011}
A.N. Norris and A.L. Shuvalov,
Elastic cloaking theory.
Wave Motion 48, 525-538 (2011).

\bibitem{farhat2009}
M. Farhat, S. Guenneau and S. Enoch, 
Ultrabroadband Elastic Cloaking in Thin Plates,
Phys. Rev. Lett. 103, 024301 (2009).

\bibitem{farhat2012}
M. Farhat, S. Guenneau and S. Enoch,
Broadband cloaking of bending waves via homogenization of multiply perforated radially symmetric and isotropic thin elastic plates.
Phys. Rev. B 85, 020301  (2012).

\bibitem{stenger2012}
N. Stenger,  M. Wilhelm and M. Wegener,
Experiments on elastic cloaking in thin plates.
Phys. Rev. Lett. 108, 014301 (2012).

\bibitem{bigoni98} D. Bigoni, S.K. Serkov, M. Valentini and Movchan, A.B.,
Asymptotic models of dilute composites with imperfectly bonded inclusions.
Int. J. Solids Struct. 35 (24), 3239-3258 (1998).

\bibitem{kim2012}
S.H. Kim and M.P. Das,  Seismic Waveguide of Metamaterials,
Mod. Phys. Lett. B 26, 1250105 (2012).

\bibitem{prl2014}
S. Br\^ul\'e, E. Javelaud, S. Enoch and S. Guenneau, Experiments on Seismic Metamaterials: Molding Surface Waves, 
Phys. Rev. Lett. 112, 133901 (2014)

\bibitem{kadic2012}
M. Kadic, T.  B\"uckmann, N.  Stenger, M. Thiel and M. Wegener, 
On the practicability of pentamode mechanical metamaterials
Appl. Phys. Lett. 100 (19), 191901, (2012)

\bibitem{kadic2013}
M. Kadic, T. B\"uckmann, R. Schittny and M. Wegener,  On anisotropic versions of three-dimensional pentamode metamaterials, New J. Phys. 15, 023029 (2013)

\end{thebibliography}
\end{document}